\documentclass[twocolumn,showkeys,showpacs,preprintnumbers,amsmath,amssymb]{revtex4}
\usepackage{graphicx}
\usepackage{textcomp}
\usepackage{epsfig}

\begin{document}

\textfloatsep 12.0pt

\def\bea{\begin{eqnarray}}
\def\eea{\end{eqnarray}}

\floatsep 30pt
\intextsep 46pt

\title{Spectral characterization of the hydrogen like atoms confined by oscillating systems.}
\author{Felix Iacob}
\email{felix@physics.uvt.ro, felix.iacob@gmail.com} 
\affiliation{{\it West University of Timi\c soara V P\^ arvan Ave 4,\\
RO 300223 Timi\c soara Romania, \\felix@physics.uvt.ro}}

\begin{abstract}

The spectral characterization of Coulomb systems confined by the homogeneous pseudo-Gaussian oscillator is investigated. This is made using the efficient computational method of generating functional. Also, the method is used for the spectral characterization of homogeneous harmonic oscillator confinement, found as a particular case of pseudo-Gaussian oscillator confinement. Finally the aspect of confinement with an impenetrable sphere of finite radius is considered by studying its conjugate effect along with harmonic oscillator upon the atom.
\end{abstract}

\keywords{Hamiltonian system; Generating functional method; Energy spectrum.}
\pacs{02.70.-c; 02.70.Hm; 03.65.Ge}

\maketitle

\section{Introduction}\label{intro}
The Coulomb systems are still of interest in literature and are used as a model in many domains of applied physics. An extensive domain, where soft-core (truncated) Coulomb systems are used to describe the interaction of intense laser fields with atoms \cite{lima}, \cite{hu}, is atomic physics. Models of hydrogen like atoms, in different confinement situations, are used to simulate the effect of high pressure on atomic static dipole polarizability. One of these models, known as hard confinement, is an impenetrable sphere of finite radius confining the atom \cite{isfr}. Thus Sommerfeld and Welkerto obtained the wave function solution in terms of confluent hypergeometric functions and emphasized that this model can be used to predict the line spectrum of hydrogen like atoms in outer atmosphere \cite{SW}.
A soft confinement of the Coulomb systems was introduced by superimposing Debye screening \cite{Ds}. This leads to the idea of using the harmonic oscillator (HO) for the atom confinement, known as the soft wall confinement (SWC) model. 
The asymptotic iteration method (AIM) was used  \cite{HSS} for the spectral characterization of a spherically confined $−a/r + br^2$ potential.
In one of his papers, studying the Zeeman effect, Avron \cite{avron} considered a Hamiltonian in which the Zeeman potential in a uniform magnetic field is replaced by the rotationally symmetric one $V \propto r^2$. Using WKB techniques he derived the large-order behavior of the perturbation series for all energy levels. The large-order behavior of SWC system was also studied by Janke and Kleinert \cite{Janke} exploiting the idea that the Coulomb system in three dimensions can be mapped onto oscillator system in four dimensions and this was made using the Bender-Wu formulas. 
Recently, the physical model of the pseudo-Gaussian oscillator (PGO) was introduced in  \cite{GF} for the Schr\" odinger and in \cite{FI} for Klein-Gordon systems, and the energy levels were determined by using the generating functional method (GFM).
The PGO potentials have HO properties approaching zero and include the genuine HO potential as limit. 
It is well known that HO potential has an infinite number of equidistant energy levels, however, PGO  potentials have  a finite number of energy levels and the interval between two consecutive levels increases slightly on higher levels.
On the other side, the GFM is a computational method developed to solve the eigenvalue problem by integration and to evaluate the expectation values for energy levels of the physical system through successive derivation.  This is an accurate method working well for systems with a finite number of energy levels.

One aspect we would like to present, in this paper, is the spectral characterization of the atom spherically confined  by the pseudo-Gaussian oscillator, using the GFM. We will say this is a finite soft wall confinement (FSWC) model.
Also, the energy spectra for SWC, found as a particular case of FSWC, is evaluated with the same technique and the obtained data values are compared with those in literature.
Another aspect we would like to present is how the energy levels values of SWC system varies by introducing the confinement of an impenetrable finite radius sphere. Also we will show that the degeneracy of energy states of free atom are lifted, in the case of confinement. 

The original inspiration, for the confinement, had come from atoms under extreme pressure and heat. Another example is the atom in solid, where the idea is that an atom confined in a Wigner-Seitz cell might develop a conduction band.
If we retain this idea, then the process by which the electron becomes ''free`` should perhaps be described as ''delocalisation`` rather than ionization. This arises when the ground state energy of the confined atom rises above the ionization threshold of free atom. Considering this,
we propose this PGO confinement model to a better understanding of both the atom under extreme pressure and heat and the metal–insulator transitions, by providing the energy values of electron in a such perturbed atom. PGO can also be used as a confinement model in the behavior of novel nano-structures such as quantum dots and quantum wires and another micro-electronic devices \cite{titus}. 
As mentioned above PGO model admits a finite number of energy levels. These number of levels can be manipulated,  by choosing a numerical value for the reference energy, $\lambda$, which sets the depth of potential well. From this point of view this PGO confined model is actually closer to the reality of these nano-devices.

Over the  time, other computational and approximate mathematical methods were developed and used to solve the eigenvalue problem of the Schr\" odinger equation with different types of potentials, which cannot be analytically solved. Let us  mention some of them: the variational method \cite{VM}, the functional analysis method \cite{FAM}, the supersymetric approach \cite{Ssym}, the asymptotic iteration method \cite{AIM}, and the factorization method \cite{FM}. Also, there are known other adapted methods to the various cases of exact quantization \cite{quant},\cite{quant1}.  

The paper is organized as follows:
 section  (\ref{pmgfm}) is dedicated to the physical model with a review upon pseudo-Gaussian potential,  in section (\ref{gfsec}) we have a brief exposure of standard GFM, section (\ref{aconfsec}) shows how the GFM works for the considered physical model and The atom confinement spectral characterization calculated results are presented in section (\ref{nrsec})
.

\section{The physical model.}\label{pmgfm}
Let us consider the radial part of the three-dimensional Schr\" odinger time-independent equation $H\psi = E\psi$. The Hamiltonian $H$, given in atomic units,
\begin{equation}\label{Ham}
H = -\frac{1}{2}\Delta + V(r),\;\;\;
\end{equation}
introduces the potential $V(r)$:
\begin{equation}\label{fswcpot}
V(r) = -\frac{1}{r} + W_{\lambda,\mu}^s(r)
\end{equation}
where, for confinement, the pseudo-Gaussian potential, $W_{\lambda,\mu}^s(r)$ is added to the hydrogen atom potential
\begin{equation}\label{W}
W_{\lambda,\mu}^s(r)=\left(\lambda+\sum_{k=1}^s C_k
r^{2k}\right)\exp(-\mu r^2)\,,
\end{equation}
with the coefficients $C_k$ defined as
\begin{equation}\label{Ck}
C_k=\frac{(\lambda+k)\mu^{k}}{k!}\,.
\end{equation}
The properties of these models are completely determined by the
dimensionless parameters $\mu,\,\lambda\in {\Bbb R}$ and the positive
integer $s=1,2,...$ which is called the order of PGO.
We note that the genuine Gaussian potential corresponding to $s=0$ is not included in this family.
The potentials defined by the eqs. (\ref{W}) and (\ref{Ck}) have the
remarkable property to approach to the potential of HO when $r\to 0$.
More, it can be proved that for each order $s$, the Taylor expansion of these potentials,
\begin{equation}\label{Tay}
W_{\lambda,\mu}^s(r)=\lambda + \mu r^2 + O(r^{2s+2})\,,
\end{equation}
does not include terms proportional with $r^4,r^6,...,r^{2s}$.
Therefore, when $s$ increases, the potential (\ref{W}) is approaching
to that of HO and the potential (\ref{fswcpot}) becomes:
\begin{equation}\label{swcpot}
V(r) = -\frac{1}{r} +\frac{1}{2}\hat \omega^2r^2
\end{equation}
with centrifugal constant $\hat \omega=\sqrt{2\mu}$ measuring the  strength of confinement.
\begin{figure}
\centering
    \includegraphics[width=6.4cm,height=5cm]{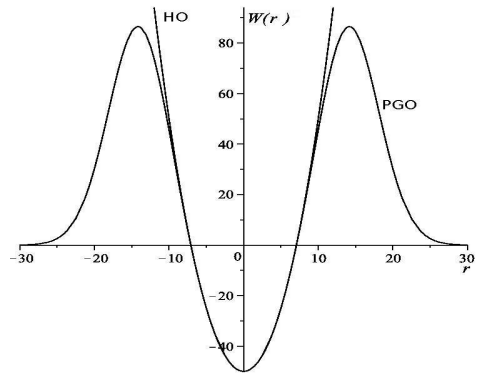}
  \caption{The pseudo-Gaussian oscillator potential graph compared with harmonic oscillator potential one.}
\label{PGO_pot}
\end{figure}
In figure (\ref{PGO_pot}) it is shown the graph of both PGO and HO to illustrates  the similar shape in a vicinity of origin.

Taking into consideration that  $l(l + 1)$ represents the eigenvalue of the square of the angular-momentum operator, the radial part of $H$ can be written explicitly as:
\begin{equation}\label{H_rad}
-{1\over 2}\left( {d^2 \over dr^2}+{2\over r} {d \over dr}-{l(l+1)\over 2r^2}\right)+V(r), 
\end{equation}
and with the potential (\ref{fswcpot}) we can study the radial three dimensional Schr\" odinger eigenvalue problem for the pseudo-Gaussian potential confinement, i.e. the FSWC model.
In the particular case of SWC, the potential (\ref{swcpot}) is used in the Hamiltonian (\ref{H_rad}), so we get the radial three dimensional Schr\" odinger eigenvalue problem for HO confinement.

\section{The generating functional method.}\label{gfsec}
To apply the GFM to this model, it is required to specify a basis for the Hilbert space of states. Considering hydrogen atom in different confinement situations, it is suitable to make use of its radial wave-functions as a canonical basis for the Hilbert space.
The radial functions, solutions of Schr\" odinger time-independent equation of the hydrogen-like atoms, are given in terms of Laguerre polynomials by:
\begin{equation}\label{RF}
R_{n,l}(r)=-\frac{2}{n^2}\sqrt{\frac{(n-l-1)!}{(n+l)!}} e^{-\frac{r}{n}} \left(\frac{2r}{n}\right)^l L_{n-l-1}^{2l+1}\left(\frac{2r}{n}\right)
\end{equation}
The generalized Laguerre polynomials  may be written with the help of generating function and according to \cite{AS} they are:
$$
L_{p}^{a}(x) = \left . \frac{1}{p!} \frac{\partial^p}{\partial \sigma^p}(1-\sigma)^{-a-1}e^\frac{x\sigma}{\sigma-1}\right |_{\sigma=0}\,,
$$
where in our case $x$ is replaced with $\left(\frac{2r}{n}\right)$ and  $p$ with $n-l-1$ respectively $a$ with $2l+1$ and $n$, $l$ are the well known, principal respectively orbital, quantum numbers.

For any radial operator ${\cal X}$ we can calculate the generating functional as:
\begin{equation}\label{intGF}
Z(\tau,\sigma)[{\cal X}] = \int_0^\infty dr\; r^2 \;R_{n^\prime,l}(r)\left[{\cal X} R_{n,l}\right](r) \,,
\end{equation}
The matrix elements of the operator ${\cal X}$, can be derived from the generating functional (\ref{intGF}) as follows:
\begin{equation}\label{rule}
\left<n^\prime\,l^\prime|\,{\cal X}\,|n\,l\right>=\delta_{l^\prime,\,l}\,\left.\frac{1}{n^\prime!\,n!} \partial_{\sigma}^{n^\prime}
\partial_{\tau}^{n}Z(\sigma,\tau)[{\cal X}]\right |_{\sigma=\tau=0}\,.
\end{equation}
For the spectral characterization ${\cal X}$ is replaced with the specific Hamiltonian operator.

The physical problem of studying the atom in different confinement situations allows us to view the Hamiltonian (\ref{H_rad}) as made up by a part describing the atom  $H_a=-\frac{1}{2}\Delta_r -\frac{1}{r}$ and one describing the confinement as a perturbative one $H_i=W_{\lambda,\mu}^s(r)$. This  approach can be found in literature, an example is Ref. \cite{CHS}.
This justifies the choice made at the beginning of this section to calculate the generating functional (\ref{intGF}) in the canonical basis of atom (\ref{RF}) using Laguerre polynomials. 
The generating functional is rewritten as:
\begin{equation}\label{gf}
Z(\tau,\sigma)[H]= Z(\tau,\sigma)[H_a] + Z(\tau,\sigma)[H_i],
\end{equation}
so the energy levels of the system are corrective to those of the atom and follows from
\begin{eqnarray}\label{Hrule}
&&\left<n^\prime\,l^\prime|\,H\,|n\,l\right>= \\\nonumber
&&\delta_{l^\prime,\,l}\,\left.\frac{1}{n^\prime!\,n!} 
\left [ \partial_{\sigma}^{n^\prime}\partial_{\tau}^{n}Z(\sigma,\tau)[H_a ]
+ \partial_{\sigma}^{n^\prime}\partial_{\tau}^{n}Z(\sigma,\tau)[H_i ] 
\right] \right |_{\sigma=\tau=0}\,
\end{eqnarray}  
In the computation process $N$ steps of derivations has been taken, so we obtain a ($N\times N$) matrix
which in general is not a diagonal one, due to the perturbative term. To obtain the values of energy levels, the matrix (\ref{Hrule}) will be subject to a diagonalization process.
The effective calculation of generating functional (\ref{gf}) is made by solving the integrals appeared in the kinetic part $Z(\tau,\sigma)[-\frac{1}{2}\Delta]$ as well as in the potential part $Z(\tau,\sigma)[r^{-1}]$ along with $Z(\tau,\sigma)[W_{\lambda,\mu}^s(r)]$.
This will be presented in the following section.

\section{The generating functional in the case of hydrogen like atoms confinement.}\label{aconfsec}

In the case of potential (\ref{fswcpot}) the GF (\ref{gf}) can be organized as a sum of terms:
\begin{equation}\label{gf_W_pow}
Z(\tau,\sigma)[H] =
        Z(\tau,\sigma)[-\frac{1}{2}\Delta] + Z(\tau,\sigma)[r^{-1}]+ Z(\tau,\sigma)[W_{\lambda,\mu}^s(r)].
\end{equation}
In the evaluation process, the advantage of this method is that the integrals (\ref{gf_W_pow}) reduce to known Gaussian ones.
To do so, we have to specify the shape of the potential, i.e. to give the order $s$ and the coefficients $\mu$ and $\lambda$. In this paper our computations are made with potentials of order $s=3$, case in which the potential (\ref{Tay}) expands:
\begin{eqnarray}\label{expand_pot}
W_{\lambda,\mu}^{s=3}(r) &=& 
\lambda + \mu r^2 +  \\\nonumber
&&\left({-\textstyle\frac{1}{6}} - {\textstyle\frac{\lambda}{24}}\right) \mu^4 r^8 + 
  \left({\textstyle\frac{1}{8}} - {\textstyle\frac{\lambda}{30}} \right) \mu^5 r^{10} + \\\nonumber
&&\left({-\textstyle\frac{1}{20}} - {\textstyle\frac{11\,\lambda}{720}}\right)  \mu^6 r^{12} + 
  \left({\textstyle\frac{1}{80}} - {\textstyle\frac{\lambda}{360}}\right)  \mu^7 r^{14} +  \ldots.
\end{eqnarray}
Having the explicit form of potential $W_{\lambda,\mu}^{s=3}(r) $, it is observed that the GF (\ref{gf_W_pow}) consists from a sum of terms containing different powers of $r$,
\begin{equation}\label{gf_pgo}
Z(\tau,\sigma)[H] =  \sum_k Z_k(\tau,\sigma)[r^{k}],\;\;k=\{-1,0,2, 8,10,12, ...\}
\end{equation}
with the calculated terms:
\begin{eqnarray}\label{gfZ}
Z(\tau,\sigma)[r^{k}] &=& 2^{2l+1}\,(n n^\prime)^{l+k+1}\\\nonumber
 &\times& \sqrt{\frac{(n-l-1)!(n^\prime-l-1)!}{(n+l)!(n^\prime+l)!}} \\\nonumber
        &\times& \frac{(2l+k+2)!\,\,[(1-\sigma)(1-\tau)]^{k+1}}{[(n+n^\prime)(1-\sigma\tau)+(n^\prime-n)(\sigma-\tau)]^{2l+k+3}},
\end{eqnarray}
where $k$ takes the values: $\{0\}$ for kinetic part, the first term in (\ref{gf_W_pow}), $\{-1\}$ for the atom part of potential, the second term in (\ref{gf_W_pow}), $\{2\}$ for HO part from pseudo-Gaussian potential (\ref{expand_pot}) and $\{8, 10, 12, \ldots\}$ for the next terms with higher power coming from pseudo-Gaussian potential expansion (\ref{expand_pot}).
In the case when $s \to \infty$, the potential (\ref{fswcpot}) tends to the potential (\ref{swcpot}) thereupon the GF (\ref{gf_pgo})
is stripped from the the higher power terms and becomes:
\begin{equation}\label{gf_ho}
Z(\tau,\sigma)[H] =
        Z(\tau,\sigma)[-\frac{1}{2}\Delta] + Z(\tau,\sigma)[r^{-1}]+ \frac{1}{2}\hat \omega^2 Z(\tau,\sigma)[r^{2}],
\end{equation}
the GF for the SWC.

\section{Numerical results}\label{nrsec}

\begin{figure}
\centering
  \vspace{0.4cm}
    \includegraphics[width=7cm,height=5cm]{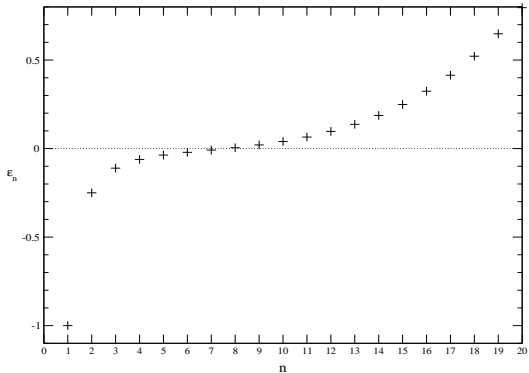}
  \caption{The calculated energies $\varepsilon_n$ of PGO confined atom, with $s=3$, $\mu=2\times 10^{-10}$ and $N=20$.}
\label{enPGO}
\end{figure}

The energy values, $\varepsilon_n$, are obtained, using the matrix elements given by:
\begin{equation}\label{Hhrule}
\varepsilon_n=\delta_{l^\prime,\,l}\,\left.\frac{1}{n!^2} 
{\cal D}iag \left [
 \partial_{\sigma}^{n}\partial_{\tau}^{n}
\sum_k Z(\tau,\sigma)[r^k]
 \right |_{\sigma=\tau=0}
\right ],
\end{equation}  
where ${\cal D}iag$ stands for the diagonalization procedure of the matrix.
First, the code was tested in the particular case of hydrogen atom where the generating functional (\ref{gf_pgo}) is:
\begin{equation}\label{gf_at}
Z(\tau,\sigma)[H_a] =
         \frac{1}{2{n^\prime}^2} \left ( \frac{1+\sigma}{1-\tau} \right )^2 
        - \left [ 1+\frac{\tau+1}{n^\prime}\,\frac{1+\sigma}{1-\tau}\right ]\,.
\end{equation}
The numerical values of energy, calculated with (\ref{Hhrule}), respect the known analytical relation $\varepsilon_n=-\frac{1}{n^2}$ so $E_n=\varepsilon_n E_0$ with $E_0=13.6\, ev$ ($n=1,2\ldots $). This confirms that the code is working properly, the values have been exactly retrieved, as they are known.
Further, we have added the pseudo-Gaussian potential, in the code, to determine the energy spectrum for the FSWC. 
The calculated numerical values of energy are presented in figure (\ref{enPGO}).
The first levels until a critical value are negative, known as bound states and the following ones are positive. In other words the energy spectra of the confined atom does not belong entirely to the negative domain, as happens in the case of the free atom. We can say that, after the ionization of the atom, the electrons can take only certain amount of energy. This seems to be reasonable in the idea of confinement, discrete spectra with positive energies were also reported in literature, one may consult \cite{isfr}, \cite{SW}.
Furthermore, it is observed an inflection point at the boundary between the negative and positive domains of energy levels. This occurs because the first energy levels are nearby to the atom nucleus and are controlled by the Coulomb part of potential. 
Moving away from the nucleus the PGO potential becomes dominant and the levels become more distant to each other.
The number of energetic values depends on the strength of confinement, so changing $\mu$ the number of negative energetic values are changed. Also, changing $s$ the number of positive energetic values are changed. 

The eigenstates with positive eigenvalues are meta-stable states.
 A particle that is bound by some attractive force is able to escape even though it lacks the energy to overcome the attractive force. Classical physics predicts that such behavior is impossible. However, the fuzziness of nature at the sub-atomic scale, that is an inherent part of quantum mechanics, implies we cannot know precisely the trajectory of α particle, this uncertainty means
the particle has a small, but non-zero probability of suddenly finding itself outside. We say it has tunneled through a potential energy barrier created by the attractive force. 
The shape of PGO potential, see fig. (\ref{PGO_pot}), appears like a well bordered by barriers. Thus, this potential admits  eigenstates with positive energies, known as meta-stable states or resonances. 

The transmission amplitude $T$ of these meta-stable states through the bordering barriers can be estimated with the help of the transfer matrix $M$ on a finite domain. An extensive presentation of transfer matrix method is exposed in \cite{TM}. 
Roughly speaking, the idea is that one can consider a barrier potential made of successive narrow constant barriers. 
In this way we express the transfer matrix $M$ as a product of matrices $M_i$. Each $M_i$  characterizes the effect of individual discontinuities of ``i-th'' sector and so the propagation through entire discretized structure is  $M=\prod_i M_i$ taken in the proper order.
The expression of transmission amplitude expressed in terms of transfer matrix  can be written as  \cite{TMC}:
\begin{equation}\label{}
T=\frac{1}{|M_{11}|^2}
\end{equation}  
A peak of the transmission amplitude does corresponds to a resonant eigenstate \cite{Res}. 
\begin{figure}
  \centering
   {\psfig{file=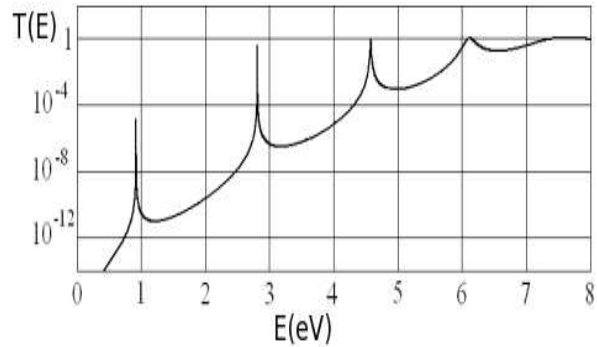,height=2.1in,width=3.2in}}
  \caption{The calculated transmission amplitude of PGO \cite{GF}. }
\label{ta}
\end{figure}
According with \cite{GF} a number of four resonant states are found for the considered potential shape ($s=3$),   corresponding to the four peaks, as figure (\ref{ta}) shows. Thus, the positive spectrum of FSWC system consists of four resonant states overlapping the continuous spectrum of energies. In an approach that computes the wave functions it is possible to predict the resonances life-time.

As the numerical data is not so accurate, by reading the figure (\ref{enPGO}), we will give the values of energy levels, expressed in atomic units, in the table (\ref{enlevcomp}). 
We observe that the values for the first energy levels, closer to nucleus, in the case of confinement do not differ very much from the first levels of free atom. The difference between these values becomes sensible just for the distant levels from nucleus. As one can observe, by confinement, the outside levels are  energetically more affected.
\begin{table*}
\caption{The  perturbed energy levels corrections of free atom with PGO and HO confinement, in (a.u.).}
\begin{tabular}{|c|c|c|c|c|}
\hline
$n$& Free atom & PGO confinement & PGO confinement  & HO confinement  \\
&& s=3 & s=18 & $\omega = 10^{-6}$\\
\hline
1 &-1.000000&-0.999994& -0.999994& -0.999997  \\
2 &-0.250000&-0.249916& -0.249916& -0.249958  \\
3 &-0.111111&-0.110697& -0.110697& -0.110904  \\
4 &-0.062500&-0.061204& -0.061204& -0.061852  \\
5 &-0.040000&-0.036850& -0.036850& -0.038425  \\
6 &-0.027778&-0.021262& -0.021262& -0.024520  \\
7 &-0.020408&-0.008354& -0.008354& -0.014381  \\
8 &-0.015625& 0.004919& 0.004919& -0.005353  \\
9 &-0.012346& 0.020540& 0.020540& 0.004097  \\
10&-0.010000& 0.040100& 0.040100& 0.015050  \\
11&-0.008264& 0.065062& 0.065062& 0.028399  \\
12&-0.006944& 0.096880& 0.096879& 0.044968  \\
13&-0.005917& 0.137057& 0.137056& 0.065570  \\
14&-0.005102& 0.187174& 0.187172& 0.091036  \\
15&-0.004444& 0.248906& 0.248897& 0.122231  \\
16&-0.003906& 0.324030& 0.323999& 0.160062  \\
17&-0.003460& 0.414433& 0.414335& 0.205487  \\
18&-0.003086& 0.522114& 0.521829& 0.259516  \\
19&-0.002770& 0.649180& 0.648406& 0.323213  \\
20&-0.002500& 0.797828& 0.795860& 0.397700  \\
\hline
\end{tabular}
\label{enlevcomp}
\end{table*}

\begin{figure}
\centering
  \vspace{0.4cm}
     \includegraphics[width=6.8cm,height=5cm]{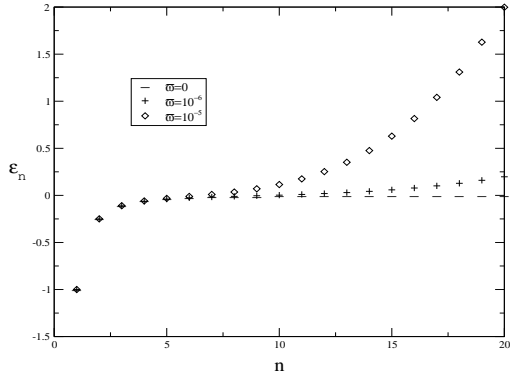}
  \caption{The calculated energies $\varepsilon_n$ for $\hat \omega=\{0,\;10^{-5},\;10^{-6}\}$.}
\label{eno}
\end{figure}

Also table (\ref{enlevcomp}) presents the calculated numerical data for HO confinement and PGO confinement with $s=18$. 
As we mentioned above, parameter $s$ measures the closeness of PGO potential to HO potential. As one can see the numerical data are closer to HO data for $s=18$ than for $s=3$, but this trend is not very pronounced and hardly visible.
To see how the atom behaves in transition from the PGO to HO confinement, calculations were made by taking higher orders of PGO. Because the values modify slightly with $s$, we will present directly the numerical results for HO confinement as the limit of PGO one, 
when $s$ equals infinity by means of a limit process.
We will use the same technique to calculate the energies $\varepsilon_n$, for this potential, having now $\hat \omega$ as a measure of confinement strength.
The behavior of energy levels for some values of $\hat \omega$ are represented graphically as a function of quantum number $n$ in figure (\ref{eno}).
Also, the numerical data values, in the case of $\hat \omega=10^{-6}$ are given in table (\ref{enlevcomp}). 
We have compared these data with those obtained by Janke and Kleinert, following the relation (33) for the energy, from their paper \cite{Janke}, we have found that our work is the case $p=2$ and $D^C=3$. The calculated perturbation coefficients, table (III) from their work,  give  energies with values comparable  to those presented here in table (\ref{enlevcomp}), HO column. However, a slightly difference between the values for higher levels is found. We think this happens due to the evaluation process of energies. 
We give the energies as calculated elements of a diagonalized matrix while they give the energies on a calculation based upon large order behavior of the perturbation series.
In figure (\ref{eno}) are also represented, by dashed line, the energy levels for the atom. This is the case of no confinement ($\hat \omega=0$) and as it is observed, the values of energy levels remain bellow zero.

\begin{figure}
 \centering
\includegraphics[scale=0.55]{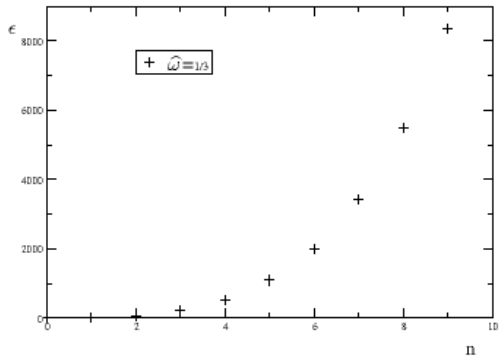}
\caption{  $\hat \omega=1/3$ means positive energy only.}
\label{enextrem}
\end{figure}

Let us see what is  the confinement strength magnitude for which the negative energies are missing. It was found that for $\hat \omega=1/3$ and above there are no negative energies. This case is presented in figure (\ref{enextrem}), and one may say there is a value for $\hat \omega$ above which the atom influence is missing.
Let us denote it with $\hat \omega_c$ and consider this is a critical value which accomplishes the condition of positive energies, $\varepsilon> 0 $ for all $\hat \omega>\hat \omega_c$. 
The numerical values of $\hat \omega_c$ are found high, for ground state, in comparison with other energy levels.
As mentioned above, for the lowest orbital $1s$ $(n=1,\; l=0$) the calculated critical value is $\hat \omega_c=1/3$. This result is in concordance with the one obtained in ref. \cite{HSS} where an analogous value of $\hat \omega_c$, denoted by $b_c$ was calculated using asymptotic iteration method and a the value $b_c=0.32533$ was found. The calculated critical values for orbitals $\{1s,\ldots 4f\}$ are presented in table (\ref{cval}).
\begin{table}
\caption{Critical values for $\hat \omega_c$ above which all energies positive.}
\begin{tabular}{|c|c|c|c|c|c|}
\hline
 orbital &4f &4d &4p &4s &3d  \\
$\hat \omega_c$ &0.000173 &  0.000125 & 0.000105 & 0.000097 & 0.000882 \\
 $b_c$& 0.00015 &0.00010 &0.00008 &0.00007 &0.00079  \\
\hline
 orbital&3p &3s &2p & 2s &1s\\ 
$\hat \omega_c$&0.000618 & 0.000537 & 0.008334 & 0.005953 & 1/3\\
 $b_c$&0.00051 &0.00042 &0.00771 &0.004831 &0.32533\\
\hline
\end{tabular}
\label{cval}
\end{table}
The range of the critical values is upper bounded by that one corresponding to orbital $1s$  and decrease, as calculation shows, for higher orbitals.

\begin{figure}
  \centering
    \includegraphics[scale=0.75]{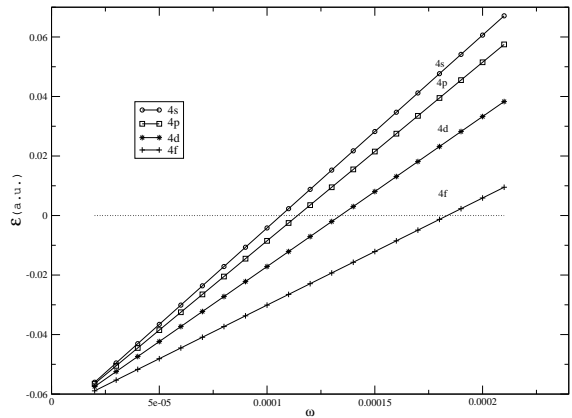}
  \caption{The degeneracy is removed as it is shown for the orbitals $4s4p4d4f$. }
\label{encritic}
\end{figure}

As a confined system, it is interesting to investigate the aspect of degeneracy of the energy. To do this, the intensity of confinement is controlled by varying $\hat \omega$, starting from the free atom ($\hat \omega = 0$) the effect of finite $\hat \omega$ is to remove the accidental degeneracy and raise the energy levels. 
The orbitals $4s\,4p\,4d\,4f$ were considered to see how degeneracy is removed with $\hat \omega$ and the results are presented in the figure (\ref{encritic}).
At small values of $\hat \omega$ the energy of the orbitals are almost identical, this is the free atom case, further increasing $\hat \omega$ with a small amount has as consequences the separation of energy levels of orbitals. Thus for fixed $n$, the higher $l$ the smaller the corresponding energy is, in other words the states get relatively less destabilized. 

\begin{figure}
\centering
  \vspace{0.4cm}
\centerline{\psfig{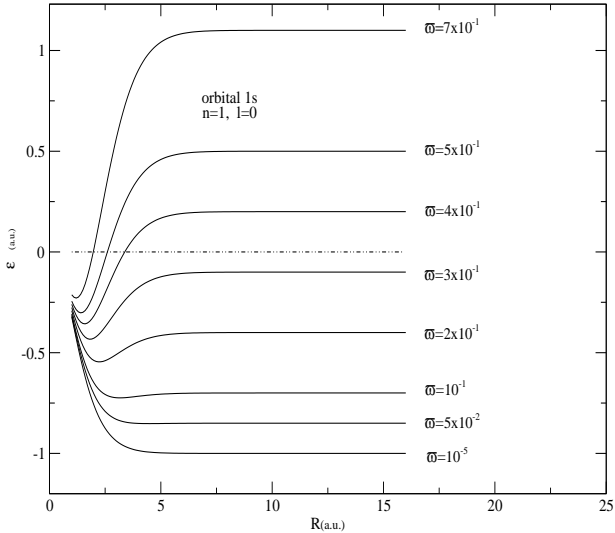}}
  \caption{Energy dependence of orbital $1s$ with both types of confinement.}
\label{Ro1s}
\end{figure}

Let us consider the confinement of an impenetrable sphere (hard  confinement) of finite radius $R$ of the system (\ref{Ham}) with the potential (\ref{swcpot}). Mathematically hard confinement means that the integrals from relation (\ref{gfZ}) will be made on a domain bounded by a sphere with radius $R$, but not smaller than one atomic radius, $R>1$. 
To see how hard  confinement affects the states of atom confined by HO system, the radius $R$ will be ranged and $\hat \omega$ will be taken as a parameter.
Our calculations have been made on the ground energy level, the orbital $1s$ and the orbital $5g$. The behavior of energy of orbital $1s$ is presented in figure (\ref{Ro1s}), considering $\hat \omega$ taking values in the set $\{ 10^{-5},\,5\times10^{-2},\,10^{-1},\,2\times10^{-1},\,3\times10^{-1},\,4\times10^{-1},\,5\times10^{-1},\,7\times10^{-1}\,\}$.
Let us notice that we can divide the domain of values of $R$ in three regions according to how confinement affects energy. 
The first region, as $R$ increases, is in a vicinity of $R=1a.u.$ where the sphere is very close to the atom. This annihilates the vibrational movement due to HO confinement, so the energy level do not depend on $\hat \omega$ and the values of energy is $\varepsilon \approx  -\frac{1}{4}a.u.$, which represents the equivalent of second energy level ($n=2$) for the free atom.
The second region follows immediately and ranges until a threshold value $R_h=5\,a.u.$, this is the region where both types of confinement are explicit.
The third region ranges beyond the threshold value, $R>R_h$, here the effect of hard confinement is weak, so the energy remains almost constant.
Let us discuss in detail what occurs in the second region.
  For relatively small $\hat \omega\approx 10^{-5}$, this is a weak SWC, it is observed that as $R$ is increased the energy goes down and stabilizes around the value of free atom, i.e. $\varepsilon \approx  -1a.u.$. This is quite well because in this conditions the atom is almost free.
 It is interesting to see the effect of conjugate action for both types of confinement for values of $\hat \omega\approx 10^{-1}$. One can see in figure (\ref{Ro1s}) that the energy as function of radius of sphere presents a minimum. This behavior does not exist for the side values of $\hat \omega\approx 10^{-5}$ and $\hat \omega\approx 7\times10^{-1}$, so
 it seems there is a resonant region, $\hat \omega\in[5\times 10^{-2},\,5\times 10^{-1}]$, where the two types of confinement, somehow, annihilate each other and the energy tends to decrease towards the free atom one.
\begin{figure}
  \centering
 \vspace{20pt}%
\centerline{\psfig{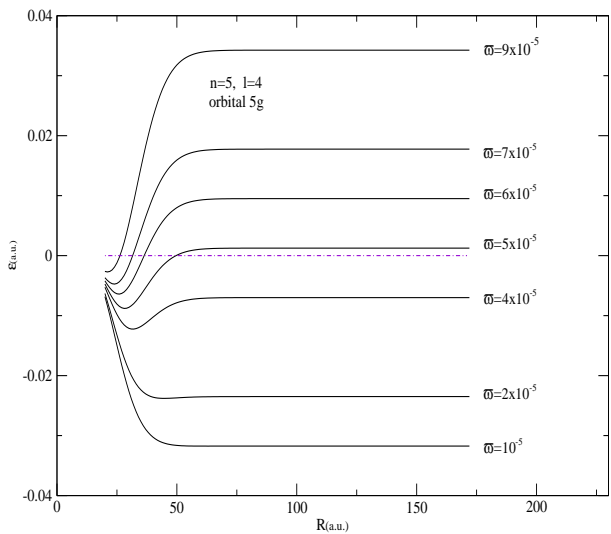}}
  \caption{Energy dependence of orbital $5g$ with both types of confinement.}
\label{Ro5g}
\end{figure}

The behavior of energy of orbital $5g$ is presented in figure (\ref{Ro5g}).
At first glance the result presented in figures (\ref{Ro1s}) and (\ref{Ro5g}) looks similar, the shape of energy as function of radius of sphere is preserved and we also have the same three regions.  A closer look indicates that the ranging domain of $R$, $\hat \omega$ and of the energy values $\varepsilon$ are different. The first region is in a vicinity of $R=25a.u$, this value is sufficient for hard confinement to annihilate the SWC for this external energy level. The resonant region of conjugate action extends up to a threshold value of $R_h=50\,a.u.$ and the value for SWC lowers to  $\hat \omega\approx 10^{-5}$. This is normal because the exterior orbitals are supposed to be affected much more by confinement than the inner ones.

It was shown, in literature, that in a confined system, a state with angular momentum $(l + 1)$ is more strongly bound than the one with $l$, which is vice versa from the aufbau principle corresponding to free atom.
\begin{figure}
  \centering
    \includegraphics[scale=1.2]{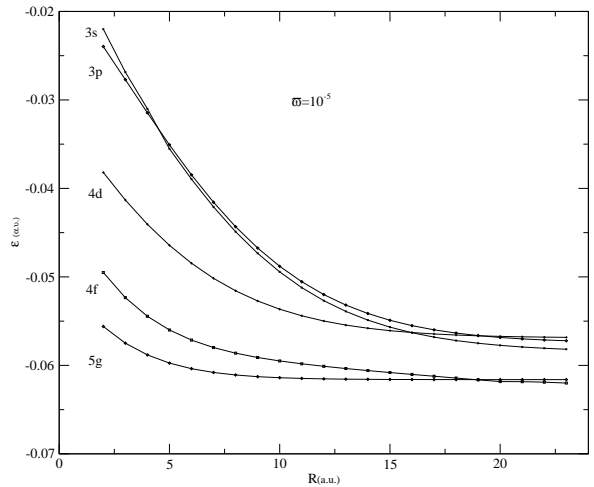}
  \caption{Crossing energy states: The level $4d$ is observed crossing $3d$ and $3s$. Also $5g$ crosses $4f$.}
\label{crossR}
\end{figure}
Taking into consideration the property of monotonicity of the range of energies, it is possible to give rise of crossing pairs of states $(n, l)$ and $ (n^\prime , l^\prime )$ with $n^\prime >n\,, l^\prime > l$. Our computation, made in the resonant region of $\hat \omega\approx 10^{-5}$, shows the existence of crossing states and  the results are presented in figure (\ref{crossR}).

In this work we have obtained novel results about the spectral characterization of perturbed hydrogen-like atoms. We have introduced PGO as an oscillating  system to confine the atom. This allows the existence of discrete positive energy spectra for the system. 
The existence of an inflection point between the negative and positive energy levels was indicated and the critical value $\hat \omega_c$  was calculated for each orbital from $1s$ to $4f$. 
It was shown that in the case of confinement the accidental degeneracy of free atom is removed and also the energy is raising as strength of confinement is increased. On the other hand, orbitals with the same $n$ are relatively less destabilized as $l$ increases. 
The case of conjugate action of hard confinement  together with HO upon the atom energy levels were studied. We have found that the energy levels as function of radius of sphere have a minimum given by a resonant action of both types of confinement.
Finally the crossing pair of energy states are calculated, they appear due to  confinement and apparently are in contradiction with the aufbau principle corresponding to the neutral free atom.

We consider this model to be useful in the explanation of the metal-insulator transition (MIT), also called the Mott transition.
An atom, found into an insulator material, is subjected to external excitations,  modeled as an oscillating system confinement and 
 may cause the atom ionization and the electron raise to the conduction band. This effect produces a spontaneous transition from insulator to metal by means of the flow of an electrical current. It was measured recently the electrical conductivity of $FeO$ as a function of pressure and temperature \cite{FeO}. Although insulating as expected at ambient condition, it was found that $FeO$ metalizes at high temperatures. Electrical conductivity of $FeO$ was measured up to $141\, GPa$ and $2480\, K$ in a laser-heated diamond-anvil cell.

\subsection*{Acknowledgments}
\noindent I would like to thank  Ion Cot\u aescu for helpful suggestions, and comments that have improved the development of this paper.

\end{document}